\documentclass[%
 reprint, 
preprint,
 amsmath,amssymb,
 aps,
pra,
]{revtex4-2}

\usepackage{graphicx}
\usepackage{dcolumn}
\usepackage{bm}
\usepackage{hyperref}
\usepackage{amsthm}
\usepackage{cancel}
\usepackage{easyReview} 

\renewcommand{\d}{{\rm d}}

\newcommand{\R}{\mathbb{R}}

\newcommand{\N}{\mathbb{N}}

\newcommand{\eq}{\begin{equation}}
\newcommand{\eqend}{\end{equation}}

\newcommand{\PD}[2]{\frac{\partial#1}{\partial#2}}
\newcommand{\DD}[2]{\frac{\d#1}{\d#2}}

\newcommand{\jv}{{\bf j}}
\newcommand{\pv}{{\bf p}}
\newcommand{\rv}{{\bf r}}

\renewcommand{\prl}{{\overleftarrow{\nabla}_{\bf r}}}
\newcommand{\prr}{{\overrightarrow{\nabla}_{\bf r}}}

\newcommand{\ppl}{{\overleftarrow{\nabla}_{\bf p}}}
\newcommand{\ppr}{{\overrightarrow{\nabla}_{\bf p}}}

\newcommand{\pPr}{{\overrightarrow{\nabla}_{\bf P}}}

\newcommand{\kv}{{\bf k}}

\newcommand{\Pv}{{\bf P}}

\newcommand{\Fv}{{\bf F}}

\newcommand{\Ev}{{\bf E}}
\newcommand{\Mv}{{\bf M}}

\newcommand{\sv}{{\bf s}}

\newcommand{\av}{{\bf a}}

\newcommand{\Av}{{\bf A}}

\newcommand{\Bv}{{\bf B}}

\newcommand{\rmi}{{\rm i}}
\newcommand{\rmd}{{\rm d}}
\newcommand{\rmw}{{\rm w}}

\newcommand{\sinc}{\,{\rm sinc}}
\newcommand{\fw}{f_\rmw}
\newcommand{\Fw}{F_\rmw}
\newcommand{\Fws}{F_{\rm s}}
\newcommand{\Trho}{\Tilde{\rho}}
\newcommand{\bro}{\Bar{\rho}}
\newcommand{\Str}{\mathcal{S}}
\newcommand{\Wig}{\mathcal{W}}
\newcommand{\Fou}{\mathcal{F}}
\newcommand{\SW}{\mathcal{T}}

\newcommand{\nr}{\nabla_\rv}
\newcommand{\ns}{\nabla_\sv}
\newcommand{\np}{\nabla_\pv}
\newcommand{\nP}{\nabla_\Pv}
\newcommand{\red}[1]{\textcolor{black}{#1}}

\newcommand{\Pcr}{\bigg(\frac{\hbar}{2}\prl\cdot\pPr\bigg)}
\newcommand{\Pcrb}{\Big(\prl\cdot\pPr\Big)}
\newcommand{\scr}{\left(\frac{\rmi\sv}{2}\cdot\nr\right)}
\newcommand{\scrb}{\bigg(\frac{\rmi\sv}{2}\cdot\nr\bigg)}
\newcommand{\scrbalt}{\bigg(\frac{\rmi}{2}\prl\cdot \sv\bigg)}

\renewcommand{\exp}[1]{\,{\rm exp}\left({ #1}\right)}
\newcommand{\expsq}[1]{\,{\rm exp}\left[{ #1}\right]}
\newcommand{\expcu}[1]{\,{\rm exp}\left\{{ #1}\right\}}
\newcommand{\expstr}{\,{ {\rm exp}\left\{\frac{1}{\rmi\hbar}\sv\cdot\left[\Pv+\frac{q}{2}\int_{-1}^1\rmd\tau\Av\left(\rv+\frac{\sv\tau}{2}\right)\right]\right\}}}
\newcommand{\expstralt}{\,{ {\rm exp}\left\{\frac{1}{\rmi\hbar}\sv\cdot\left[\Pv+q\sinc\left(\frac{\rmi\sv}{2}\cdot\nr\right)\Av(\rv)\right]\right\}}}
\usepackage{tikz}
\newcommand{\hcancel}[1]{
    \tikz[baseline]{
        \node[inner sep=0pt,outer sep=0pt] (eq) {\(\displaystyle #1\)};
        \draw (eq.south west) -- (eq.north east);
    }}
\newtheorem{definition}{Definition}[section]
\newtheorem{lemma}[definition]{Lemma}
\newtheorem{theorem}[definition]{Theorem}

\begin{document}

\preprint{APS/123-QED}

\title{Gauge-invariant Wigner equation for electromagnetic fields:\\Strong and weak formulation}

\author{Clemens Etl}
    \email{clemens.etl@tuwien.ac.at}
\author{Mauro Ballicchia}%
    \email{mauro.ballicchia@tuwien.ac.at}
\author{Mihail Nedjalkov}%
    \email{mihail.nedialkov@tuwien.ac.at}
\author{Hans Kosina}%
    \email{hans.kosina@tuwien.ac.at}
\affiliation{%
 Institute for Microelectronics, TU Wien, Gu{\ss}hausstraße 27-29/E360, 1040 Vienna, Austria
}%

\date{\today}

\begin{abstract}
Gauge-invariant Wigner theory describes the quantum-mechanical evolution of charged particles in the presence of an electromagnetic field in phase space, which is spanned by position and kinetic momentum. This approach is independent of the chosen potentials, as it depends only on the electric and magnetic field variables. 
Several approaches to derive a gauge-invariant Wigner evolution equation have been reported, which are generally complex.
This work presents a new formulation for a single electron in a general electromagnetic field based solely on differential operators that simplify existing formulations. A gauge-dependent equation is derived first using Moyal's equation. A transformation of the Wigner function, introduced by Stratonovich, is then used to make the equation gauge-invariant, which gives us a strong formulation of the problem. This equation can be transformed into its weak form, which proves that both formulations are equivalent. An analysis of the different properties of the gauge-dependent and gauge-invariant formulations is given, as well as the different requirements for the regularity and asymptotic behavior of the strong and weak formulations.
\end{abstract}

\keywords{Single electron dynamics, quantum electron transport, Wigner theory, gauge-invariance}
\maketitle


\begin{widetext}

\section{Introduction}
The quantum evolution of a charged particle in an electromagnetic (EM) field  \cite{arnold1989electromagnetic, chang2008berry, PhysRevB.96.144303, Weinbub2018}, which is relevant to areas such as nanoelectronics \cite{kluksdahl1988quantum, Ferry_and_Shifren, querlioz2006improved, jacoboni2010theory, nedjalkov2011wigner, duque2012response, bookStochastic, Cepellotti2021, ibarra2022dirac, magro2023deterministic, xue2024si}, can be described using various formulations, including gauge-invariant quantum mechanics~\cite{Serimaa86, Serimaa87, nedjalkov2022gauge}, Wigner theory~\cite{wigner1932}, and phase space trajectories governed by Fredholm integral equations~\cite{etl2023wigner}. 
Such theoretical frameworks are also fundamental to the description of entanglement~\cite{ballicchia2025approximate} and thus highly relevant for modern applications in quantum cryptography~\cite{casado2008wigner} and correlated systems involving Coulomb interactions~\cite{benam2021computational}.
The most common description of a quantum mechanical system is given by the Schr\"{o}digner equation, where the particle's state is represented by the wave function $\psi$. 
It is generalized for mixed states by the von Neumann equation for the density matrix $\rho$. Both approaches yield the probability density of the particle's location. 
To evaluate the probability density of the momentum, these functions can be transformed by the unitary Fourier transform involving momentum variables.

The Wigner formalism assesses the density in phase space using the location and momentum of a particle. The Wigner function is introduced by applying a Weyl transform 
to the density matrix $\rho$,
\begin{equation}
    \fw(\pv,\rv)=C\int \rmd\sv\exp{\frac{1}{\rmi\hbar}\sv\cdot\pv}\rho\left(\rv+\frac{\sv}{2},\rv-\frac{\sv}{2}\right),
     \label{eq:wigner}
\end{equation}
where a normalization constant $C=(2\pi\hbar)^{-3}$ is introduced.
As $\fw$ is normalized but not non-negative, it represents a quasi-probability density in phase space. The foundation for describing the evolution of $\fw$ was laid by Wigner, who transferred the density matrix using a Weyl transform, which consists of a coordinate transform followed by a Fourier transform.
However, Wigner's work did not include the equation that describes the evolution of a quantum mechanical system for a general potential $V$.
Groenewold \cite{groenewold1946principles} investigated the mathematical structure of the Wigner function and introduced the $\star$-product 
\begin{equation}
    \star\equiv\expsq{\frac{\rmi\hbar}{2}(\prl\cdot\ppr-\ppl\cdot\prr)},
\end{equation}
which was essential for fully describing the dynamics of the Wigner function. Moyal used the $\star$-product in his work \cite{Moyal49} to formulate the Wigner equation (WE) for a given Hamiltonian $H$: 
\begin{equation}
    \PD{\fw}{t}=\{H,\fw\}.
\end{equation}
The Moyal bracket is defined as
\begin{equation}
    \{H,\fw\}:=H(\pv,\rv,t)\frac{2}{\hbar}{\rm sin}\bigg[\frac{\hbar}{2}(\prl\cdot\ppr-\ppl\cdot\prr)\bigg]\fw(\pv,\rv,t),
    \label{eq:moyal}
\end{equation}
where the sine function is used in the form of its series expansion, and the arrows denote the differential operator's direction of action. In the classical limit $\hbar\rightarrow 0$, the Moyal bracket reduces to the Poisson bracket, which establishes the connection to classical mechanics.

The Weyl transform introduces the canonical momentum $\pv$ as a new variable which, in the case of an EM field, depends on the choice of the electrodynamic potentials $(\Av,\phi)$. 
This means that for a given problem, we have different formulations and thus different solutions. 
This can be avoided by replacing $\pv$ with the kinetic momentum $\Pv=\pv-q\Av(\rv)$, as it is an observable and therefore uniquely defined for a given problem. Few approaches have been made to derive a formulation of a gauge-invariant Wigner equation (GIWE)~\cite{KuboWmag64, Serimaa86, Serimaa87,perepelkin2024wigner, xavier2024generalized}. 

One way to modify $\fw$ so that it becomes gauge-invariant by using the vector potential was given by Stratonovich \cite{stratonovich1956gauge, PRBMagnetic, nedjalkov2022gauge}. 
For this purpose, a gauge-invariant density matrix $\bro$ is introduced,
\begin{align}
    \bro(\sv,\rv)&= {\rm exp}\biggl[-\frac{\rmi q}{2\hbar}\int_{-1}^{1} 
        \Av\left(\rv+\frac{\sv\tau}{2}\right)\cdot\sv\;\rmd\tau\biggr]\Trho(\sv,\rv),
\end{align}
where \begin{align}
        \Trho(\sv,\rv)&=\rho\Big(\rv+\frac{\sv}{2},\rv-\frac{\sv}{2}\Big).
\end{align}
The Weyl-Stratonovich (WS) function $\Fws$ is defined as the Fourier transform of $\bro$ with respect to the variable $\sv$,
\begin{equation}
    \Fws(\Pv,\rv):=C\int\d\sv\,{\rm exp}\bigg(\frac{1}{\rmi\hbar}\sv\cdot\Pv\bigg)\bro(\sv,\rv).
    \label{eq:WSfunction}
\end{equation}
In \cite{PRBMagnetic}, the GIWE was derived directly from the von Neumann equation, whereas in this work we use  Moyal's formulation of the WE, like in the historical evolution of the standard Wigner theory, described around \eqref{eq:wigner}--\eqref{eq:moyal}. This leads to a strong formulation in which the interaction with the EM field is represented by differential operators. 
We will prove the equivalence of the weak and strong formulations and investigate the specific requirements of the physical quantities involved. 



In the following section, we will derive a gauge-dependent formulation of the WE (GDWE) using Moyal's approach, which is then transformed into a weak form using integrals. In section~\ref{strat}, the Weyl-Stratonovich transform is introduced, where we prove its gauge-invariance and show the properties needed for deriving the GIWE. In section~\ref{giwe}, the GDWE is used to obtain both the strong and weak formulation of the problem. The results are summarized in section~\ref{conclusion}.

\section{Gauge-dependent Wigner equations}

The Hamiltonian for a charged particle with mass $m$ and charge $q$ in a general EM field is given by
\begin{equation}
    H(\pv,\rv)=\frac{1}{2m}[\pv-q\Av(\rv)]^2+q\phi(\rv).
    \label{eq:hamiltonian}
\end{equation}
The EM field is represented by the electrodynamic potentials $\Av$ and $\phi$ as
\begin{equation}
    \Bv=\nabla\times\Av,\quad \Ev=-\nabla\phi-\PD{\Av}{t}.
\end{equation}
The terms in \eqref{eq:hamiltonian} that do not depend on the momentum $\pv$ are summarized in the modified potential
\begin{equation}
    \Tilde{V}(\rv):=q\phi(\rv)+\frac{q^2}{2m}\Av^2(\rv),
    \label{eq:V}
\end{equation}
so that $H$ can be written as
\begin{equation}
    H(\pv,\rv)=\frac{\pv^2}{2m}-\frac{q}{m}\pv\cdot\Av(\rv)+\Tilde{V}(\rv).
    \label{eq:hamiltonian2}
\end{equation}
We will insert the Hamiltonian \eqref{eq:hamiltonian2} into~\eqref{eq:moyal} to derive the strong formulation of the GIWE.
Equation \eqref{eq:moyal} can be reformulated by applying the angle subtraction formula for the sine.
\begin{multline}
    \sin\bigg[\frac{\hbar}{2}\Big(\prl\cdot\ppr-\ppl\cdot\prr\Big)\bigg] = 
    \sin\bigg(\frac{\hbar}{2}\prl\cdot\ppr\bigg)\cos\bigg(\frac{\hbar}{2}\ppl\cdot\prr\bigg) \\
    -\cos\bigg(\frac{\hbar}{2}\prl\cdot\ppr\bigg)\sin\bigg(\frac{\hbar}{2}\ppl\cdot\prr\bigg)
    \label{eq:subFormula}
\end{multline}
The lowest terms of the series expansions on sine and cosine are given as
\begin{align}
    \cos\bigg(\frac{\hbar}{2}\ppl\cdot\prr\bigg)&=1-\frac{\hbar^2}{8}\Big(\ppl\cdot\prr\Big)^{2}+\dots,\label{eq:cosexp}\\
    \sin\bigg(\frac{\hbar}{2}\ppl\cdot\prr\bigg)&=\frac{\hbar}{2}\ppl\cdot\prr-\frac{\hbar^3}{48}\bigg(\ppl\cdot\prr\bigg)^3+\dots.\label{eq:sinexp}
\end{align}
As $\Tilde{V}$ is independent of $\pv$, the application of $\ppl$ in \eqref{eq:cosexp} and \eqref{eq:sinexp} yields zero. Thus, only the $1$ in the cosine gives a non-zero contribution, so that we can write
\begin{equation}
    \Tilde{V}\sin\bigg[\frac{\hbar}{2}(\prl\cdot\ppr-\ppl\cdot\prr)\bigg]=\Tilde{V}\sin\bigg(\frac{\hbar}{2}\prl\cdot\ppr\bigg).
    \label{eq:Vtilde}
\end{equation}
As $\pv\cdot\Av$ is linear in $\pv$, the first summand of both the sine and cosine expansion yields a non-zero result, which gives us
\begin{align}
    \pv\cdot\Av\sin\bigg[\frac{\hbar}{2}(\prl\cdot\ppr-\ppl\cdot\prr)\bigg] 
    &=\pv\cdot\Av\sin\bigg(\frac{\hbar}{2}\prl\cdot\ppr\bigg)-\frac{\hbar}{2}\Av\cdot\prr\cos\bigg(\frac{\hbar}{2}\prl\cdot\ppr\bigg),\\
    \pv^2\sin\bigg[\frac{\hbar}{2}(\prl\cdot\ppr-\ppl\cdot\prr)\bigg] 
    &=\pv^2\sin\bigg(\frac{\hbar}{2}\prl\cdot\ppr\bigg)-\hbar\pv\cdot\prr\cos\bigg(\frac{\hbar}{2}\prl\cdot\ppr\bigg),
    \label{eq:psquared}
\end{align}
for the first two terms of \eqref{eq:hamiltonian2}, where \eqref{eq:psquared} contains terms that are zero that help us to keep the equation compact. By inserting \eqref{eq:hamiltonian2} into \eqref{eq:moyal} and using \eqref{eq:Vtilde}--\eqref{eq:psquared} we get
    \begin{theorem}[Strong formulation of the GDWE]
    Let $\Trho_0$ be the density matrix at $t=0$ in an EM field with smooth electrodynamic potentials $(\Av,\phi)$. For the Wigner function with the initial condition
    \begin{equation}
        f_{\rmw}^0(\pv,\rv)=C\int\rmd\sv\exp{\frac{1}{\rmi\hbar}\sv\cdot\pv}\Trho_0(\sv,\rv),
    \end{equation}
    the evolution of $\fw$ is given by
    \begin{align}
        \PD{\fw}{t}=\bigg[\frac{1}{\hbar m}(\pv-q\Av)^2+\frac{2q}{\hbar}\phi\bigg]\sin\bigg(\frac{\hbar}{2}\prl\cdot\ppr\bigg)\fw 
        -\frac{1}{m}(\pv-q\Av)\cdot\cos\bigg(\frac{\hbar}{2}\prl\cdot\ppr\bigg)\nr\fw.
        \label{eq:GDWEstrong}
    \end{align}
\end{theorem}
\red{Interestingly, the kinetic momentum $\pv-q\Av$ appears twice in this equation. The second term on the right-hand side resembles a diffusion term.}
The potentials $\Av$ and $\phi$ must be smooth in the given form.

To achieve gauge-invariance for the WE, a simple variable transform $\pv\mapsto\Pv+q\Av(\rv)$ in \eqref{eq:GDWEstrong} does not suffice, because the additional term $q\Av(\rv)$ does not compensate for the change of the density matrix under gauge transforms.
Stratonovich's approach \cite{stratonovich1956gauge} to solving this issue was a modification of the Weyl transform. In the next section, we will explore this transformation and show its necessity for deriving the GIWE.

\section{Weyl-Stratonovich transform}
\label{strat}

A gauge transform for given potentials $(\Av,\phi)$ can be expressed using a scalar function $\chi$, so that the new potentials
\begin{equation}
    \Av'=\Av+\nabla\chi,\quad\phi'=\phi-\PD{\chi}{t}
\end{equation}
represent the same EM field:
\begin{equation}
    \Bv=\nabla\times\Av,\quad\Ev=-\nabla\phi-\PD{\Av}{t}.
\end{equation}
The density matrix $\Trho'$, which depends on the gauge, can be evaluated using $\Trho$ of another gauge as
\begin{equation}
    \Trho'(\sv,\rv,t)=\expcu{-\frac{1}{\rmi\hbar} q\Big[\chi\Big(\rv+\frac{\sv}{2},t\Big)-\chi\Big(\rv-\frac{\sv}{2},t\Big)\Big]}\Trho(\sv,\rv,t).
\end{equation}
This means that when the Wigner function is expressed in terms of the kinetic momentum $\Pv$, i.e., $\Fw(\Pv,\rv,t):=\fw(\Pv+q\Av,\rv,t)$, it remains gauge-dependent, see Appendix~\ref{eq:gdFw}.
To remove the gauge-dependence, we use Stratonovich's approach, which we introduce in the following transformation.
\begin{definition}[Weyl-Stratonovich (WS) transform]
    Let $\Av$ be a vector potential and $f\in L^1(\R^3\times\R^3)$. The WS transform is defined as
    \begin{equation}
        \Str:f\mapsto C\int\rmd\sv\expstr f(\sv,\rv).
    \end{equation}
The Stratonovich function is defined as the WS transform of the density matrix $\Trho(\sv,\rv)$, i.e., 
\begin{align}
    \Fws(\Pv,\rv):=
    C\int\rmd\sv\expstr\Trho(\sv,\rv).
\end{align}
\end{definition}
For $\Fws$ we can show
\begin{lemma}
    Let $\Av$ be a vector potential and $\rho$ a solution of the von Neumann equation. Then, the Stratonovich function $\Fws$ is invariant under gauge transformations.
    \label{lemmaGIFs}
\end{lemma}
For the proof, check Appendix~\ref{proofGIFs}.
When we replace all $\fw$  by $\Fws$ in \eqref{eq:GDWEstrong}, we eliminate the potentials $(\Av,\phi)$ and obtain an equation that contains only the fields $\Ev$ and $\Bv$. For this purpose, $\fw$ is transformed back to the density matrix using an inverse Weyl transform
\begin{equation}
    \Wig^{-1}: \fw(\pv,\rv)\mapsto\int\rmd\pv\exp{-\frac{1}{\rmi\hbar}\sv\cdot\pv}\fw(\pv,\rv)\quad\Big(=\Trho(\sv,\rv)\Big),
\end{equation}
followed by a WS transform. This concatenation yields a transformation $\SW:=\Str\circ\Wig^{-1}$ that satisfies
\begin{equation}
    \SW(\fw)=\Fws.
\end{equation}
We will apply this transformation to both sides of \eqref{eq:GDWEstrong}, but first, we investigate its properties in more detail.

The linearity of $\SW$ follows directly from the linearity of the Fourier transform
\begin{equation}
    \SW(a\fw+bg_\rmw)=a\Fws+bG_{\rm s},\label{eq:Clin}
\end{equation}
where $a$ and $b$ can be functions of $\rv$ and $t$. The transformation of the differential operators is given by
\begin{align}
    \SW\bigg(\PD{\fw}{t}\bigg)&=\PD{\Fws}{t}-q\PD{\Av}{t}\cdot\sinc\Pcr\nP\Fws,\label{eq:fwdt}\\
    \SW(\nr\fw)&=\nr\Fws-q\nr(\pPr\cdot\Av)\sinc\Pcr\Fws,\label{eq:fwdr}\\
    \SW(\np\fw)&=\nP\Fws,\label{eq:fwdp}
\end{align}
which is shown in Appendix~\ref{Cderivative}. \red{Eq.~\eqref{eq:fwdt} shows an important difference between $\fw$ and $\Fws$ with respect to stationary states. While the condition $\partial\fw/\partial t=0$ has different solutions for different choices of $\Av$, the solutions of $\partial\Fws/\partial t=0$ remain the same. Only when $\Av$ is time-independent are the stationary states for both functions the same.}

The term $\pv\fw$ transforms as
\begin{align}
    \SW(\pv\fw)=\bigg[&\Pv+\frac{q\hbar}{2}\nr\Big(\pPr\cdot\Av\Big)\sinc'\Pcr 
    +q\Av\sinc\Pcr\bigg]\Fws,\label{eq:pfw}
\end{align}
see Appendix~\ref{CP}.

We apply $\SW$ to both sides of the gauge-dependent WE \eqref{eq:GDWEstrong} and use the properties \eqref{eq:Clin}--\eqref{eq:pfw} to derive the strong formulation of the GIWE.
\section{Gauge-invariant Wigner equations}
\label{giwe}
Since the derivation is elaborate and contains many terms, we discuss only the steps in this section. The detailed derivation of each step is given in Appendix~\ref{DevSF}.
\subsection{Strong formulation}
\begin{enumerate}
    \item The linearity of $\SW$ enables us to transform each term in \eqref{eq:GDWEstrong} one by one. 
    We use \eqref{eq:fwdt}--\eqref{eq:pfw} to transform all $\fw$'s into $\Fws$'. In the remaining steps, all terms containing $\Av$ and $\phi$ are replaced by terms containing $\Ev$ and $\Bv$.
    \item 
    We derive the identity
    \begin{align}
        \av\times\Bv&=\av\times(\nr\times\Av)\nonumber\\
        &=\nr(\av\cdot\Av)-(\av\cdot\nr)\Av,\label{eq:vtp}
    \end{align}
    which we will use later, where $\av$ stands for $\Pv$ or $\pPr$ respectively.
    \item One term contains $\Av^2$, which we have to split up to be able to make use of the identity \eqref{eq:vtp}. We use the addition formula to rewrite it as
    \begin{align}
        \Av^2\sin\Pcr 
        &=\Av'\cdot\Av''\sin\bigg[\frac{\hbar}{2}\Big(\prl'+\prl''\Big)\cdot\pPr\bigg]\nonumber\\
        &=2\bigg[\Av\sin\Pcr\bigg]\cdot\bigg[\Av\cos\Pcr\bigg],
        \label{eq:Asquared}
    \end{align}
    where the product rule was considered by marking $\Av$ and $\prl$ with $'$ and $''$ to identify which $\nabla$-operator acts on which vector potential.
    \item We also want all of the pseudo-differential operators to be either of type $\sinc$ or $\sinc'$. Therefore, we use
    \begin{align}
        \sin(x)&=x\sinc(x),\label{eq:sinSinc}\\
        \cos(x)&=\sin'(x)\nonumber\\
        &=[x\sinc(x)]'\nonumber\\
        &=\sinc(x)+x\sinc'(x).\label{eq:cosSinc}
    \end{align}
    to replace the sines and cosines, where we set $x=\hbar/2\prl\cdot\pPr$.
    \item Steps 1 to 4 yield many terms, where some of them cancel each other. The term containing $\partial\Av/\partial t$, which arises through the transformation of $\partial\fw/\partial t$, can be combined with $\nr\phi$ so that we can use $\Ev=-\nr\phi-\partial\Av/\partial t$. For the remaining terms, we use the identity \eqref{eq:vtp}. In this way, we can indeed eliminate all potentials.
    \item Simplifying the remaining terms and using the definition of the Lorentz force $\Fv=q\Ev+qm^{-1}\Pv\times\Bv$ yields 
\end{enumerate}
\begin{theorem}[Strong formulation of the GIWE]
    \label{strongGIWE}
    Let $\Trho_0$ be the density matrix at $t=0$ in an EM field $(\Ev, \Bv)$. For the Stratonovich function with the initial condition
    \begin{align}
        \Fws^0(\Pv,\rv)=C\int\rmd\sv\expstr
        \Trho_0(\sv,\rv),
    \end{align}
    the evolution of $\Fws$ is given by
    \begin{multline}        
    \PD{\Fws}{t}+\frac{\Pv}{m}\cdot\nr\Fws+\Fv\cdot\sinc\Pcr\nP\Fws 
        = \\
        -\frac{q\hbar}{2m}\Big(\pPr\times\Bv\Big)\sinc'\Pcr
        \cdot\bigg[\prr-q\Big(\pPr\times\Bv\Big)\sinc\Pcr\bigg]\Fws.
        \label{eq:WEstrong}
    \end{multline}
\end{theorem}
In the classical limit $\hbar\to 0$, the right-hand side vanishes, and the $\sinc$ function on the left-hand side becomes 1, which gives us the classical Liouville equation
\begin{equation}      
    \PD{\Fws}{t}+\frac{\Pv}{m}\cdot\nr\Fws+\Fv\cdot\nP\Fws= 0.
\end{equation}

The term containing $\nr\Fws$ can be interpreted as a generalized diffusion term:
\begin{equation}
    \bigg[\frac{\Pv}{m}+\frac{q\hbar}{2m}\Big(\pPr\times\Bv\Big)\sinc'\Pcr\bigg]\cdot\nr\Fws.
\end{equation}
In \eqref{eq:diffT}, one can see that the transformation of the diffusion term of $\fw$ introduces two terms that do not contain $\nr\Fws$. 
This happens due to the transformation \eqref{eq:fwdr} of $\nr$, where a second term is introduced that contains $\nr(\pPr\cdot\Av)$. 
I.e., the diffusion term of $\fw$ is gauge-dependent and only equal to the  diffusion term of $\Fws$ when $\Av$ is spatially constant, which is equivalent to having no magnetic field. 
This enables the evaluation of the particle diffusion, which is not provided for the gauge-dependent formulation. 

The right-hand side of \eqref{eq:WEstrong} consists of a linear term in $\Bv$ and
a quadratic term in $\Bv$. As far as magnetism is concerned, the linear term describes the paramagnetic effect and the quadratic term the diamagnetic effect. Depending on the material under consideration, the quadratic term can be either negligible or dominant. 
When modeling light-matter interaction, in many practical cases, it is sufficient to keep the linear interaction term. The widely used electric dipole and quadrupole interaction operators are linear in the field variables.


In \cite{Levanda01} a gauge-invariant evolution equation was derived for the same case of a single electron in an EM field but a different Wigner quasi-distribution function $\rho_\rmw$. If we use the same notation, we can write the equation as
\begin{equation}
    \Big(\rlap{/}\partial^T+\frac{1}{m}\rlap{/}\Pv\cdot\rlap{\hspace{0.5mm}/}\nabla^X\Big)\Fws=0,
\end{equation}
where
\begin{align}
    \rlap{/}\partial^T&=\PD{}{t}+q\Ev\cdot\pPr\sinc\Pcr,\\
    \rlap{/}\Pv&=\Pv+\frac{\hbar q}{2}\Big(\pPr\times\Bv\Big)\sinc'\Pcr,\\
    \rlap{\hspace{0.5mm}/}\nabla^X&=\prr-q\Big(\pPr\times\Bv\Big)\sinc\Pcr.
\end{align}
In this case, the right-hand side of the equation vanishes in contrast to equation (31) in \cite{Levanda01}, where there is an exponential function that includes derivatives of time, energy, momentum, and location. This simplification becomes obvious when we consider how $\rho_\rmw$ in \cite{Levanda01}, which uses four-vectors, is related to the Stratonovich function. The function $\rho_\rmw$ uses another variable $\epsilon$, which represents the total energy of the electron. It can be shown that $\rho_\rmw$ can be transformed into $\Fws$ by integrating over $\epsilon$, see Appendix~\ref{rhowVSfs}. The Stratonovich function can thus be seen as a marginal distribution of $\rho_\rmw$, which explains the similar, simplified structure of the strong formulation of the GIWE. This construction resembles the relation between the non-equilibrium Green's function and the density matrix in quantum transport theory, where the energy-integrated Green's function yields the particle density in phase space~\cite{wang2014nonequilibrium}.

Next, we will transform this equation into its integral form to establish a connection to the GIWE derived in \cite{PRBMagnetic}, and compare the requirements for the EM field regarding regularity and asymptotic behavior.
\subsection{Weak formulation}
Before we can transform the strong formulation of the GIWE into the weak formulation, we first need to investigate the pseudo-differential operators to replace the nabla operator of $\Pv$ with $\sv$. This can be achieved by using the following lemmas. Consider that the Fourier transform $\Fou$ of a function $f(\sv)$ actually gives us a function that depends on the wave vector $\kv$, but since we use $\Pv$, we define the Fourier transform as
\begin{equation}
    \Fou(f)(\Pv):=\sqrt{C}\int\d\sv\,{\rm exp}\bigg(\frac{1}{\rmi\hbar}\Pv\cdot\sv\bigg)f(\sv).
\end{equation}
\begin{lemma}
    Let $g\in C^\infty(\R)$ be an analytic function and $f\in L^1(\Omega),\Omega\subseteq\R^d$.
    For the Fourier transform $\Fou:f(\sv)\mapsto\hat f(\Pv)$ holds
    \begin{equation}
        \Fou\Big[g\Big(\rmi\prl\cdot\sv\Big)f(\sv)\Big]=g\Big(-\hbar\prl\cdot\pPr\Big)\Fou[f(\sv)].
    \end{equation}
    \label{stonp}
\end{lemma}
The proof can be found in Appendix~\ref{proofStonp}. When we set $g$ to $\sinc$ and $\sinc'$, we get
\begin{align}
    \sinc\bigg(\frac{\hbar}{2}\prl\cdot\pPr\bigg)\Fou(f)&=\Fou\bigg[\sinc\scrbalt f\bigg].\label{eq:FouSin}\\
    \sinc'\bigg(\frac{\hbar}{2}\prl\cdot\pPr\bigg)\Fou(f)&=-\Fou\bigg[\sinc'\scrbalt f\bigg],\label{eq:FouCos}    
\end{align}
By investigating the expansion of $\sin$ and $\cos$ on the right-hand side of \eqref{eq:FouCos}--\eqref{eq:FouSin}, we can use a different representation, given in the following Lemma.
\begin{lemma}
    Let $g\in C^\infty(\R^n), n\in\N$. Then, there holds
    \begin{align}
        g(\rv)\sinc\scrbalt&=\frac{1}{2}\int_{-1}^1\rmd\tau g\bigg(\rv+\frac{\sv\tau}{2}\bigg),\label{eq:sincInt}\\
        g(\rv)\sinc'\scrbalt&=-\frac{\rmi}{2}\int_{-1}^1\rmd\tau g\bigg(\rv+\frac{\sv\tau}{2}\bigg)\tau.\label{eq:sincdiff}
    \end{align}
    \label{trigLemma}
\end{lemma}
The proof is shown in Appendix~\ref{trigLemmaProof}. 
If we fix $\rv$ in the density matrix, i.e., $\Trho_\rv(\sv):=\Trho(\sv,\rv)$, we can interpret the SF as a Fourier transform, where
\begin{equation}
    \Fws=\sqrt C\Fou\bigg\{{\rm exp}\bigg[\frac{q}{2\rmi\hbar}\sv\cdot\int_{-1}^1\d\tau\Av\Big(\rv+\frac{\sv\tau}{2}\Big)\bigg]\Trho_\rv\bigg\}.\label{eq:fouRho}
\end{equation}
This allows for using the properties of the Fourier transform, like the convolution theorem $\Fou(u)*\Fou(v)=\sqrt{C^{-1}}\Fou(u\cdot v)$, which we apply to the third term on the left and the expression on the right-hand side of \eqref{eq:WEstrong}. For this purpose, they have to be expressed in the form of $\Fou(u\cdot v)$. According to \eqref{eq:fouRho}, we set $v=\sqrt C{\rm exp}[q(2\rmi\hbar)^{-1}\sv\cdot\int_{-1}^1\d\tau\Av(\rv+\sv\tau/2)]\Trho_\rv$ so that $\Fou(v)=\Fws$. To figure out $u$ for the third term on the left of \eqref{eq:WEstrong}, we use Lemma~\ref{stonp} and Lemma~\ref{trigLemma} to show that
\begin{align}
    \Fou(u\cdot v)&=\Fv\cdot\sinc\Pcr\nP\Fws\nonumber\\
    &=\Fv\cdot\sinc\Pcr\nP\Fou(v)\nonumber\\
    &=\Fou\bigg[\frac{1}{\rmi\hbar}\Fv\cdot\sv\sinc\scrbalt v\bigg]\nonumber\\
    &=\Fou\bigg[\frac{1}{2\rmi\hbar}\int_{-1}^1\rmd\tau \Fv\bigg(\rv+\frac{\sv\tau}{2}\bigg)\cdot\sv v\bigg],\nonumber\\
    \Rightarrow u(\sv)&=\frac{1}{2\rmi\hbar}\int_{-1}^1\rmd\tau \Fv\bigg(\rv+\frac{\sv\tau}{2}\bigg)\cdot\sv.\label{eq:uGI1}
\end{align}
Using \eqref{eq:sincdiff} for the right-hand side of \eqref{eq:WEstrong} yields
\begin{align}
    \Fou(u\cdot v)&=\Big(\pPr\times\Bv\Big)\cdot\prr\sinc'\Pcr\Fou(v)\nonumber\\
    &=\Fou\bigg[-\frac{1}{\rmi\hbar}\Big(\sv\times\Bv\Big)\cdot\prr\sinc'\scrbalt v\bigg]\nonumber\\
    &=\Fou\bigg[\frac{1}{2\hbar}\int_{-1}^1\rmd\tau\bigg[\sv\times\Bv\bigg(\rv+\frac{\sv\tau}{2}\bigg)\bigg]\tau\cdot\nr v\bigg],\nonumber\\
    \Rightarrow u(\sv)&=\frac{1}{2\hbar}\int_{-1}^1\rmd\tau\bigg[\sv\times\Bv\bigg(\rv+\frac{\sv\tau}{2}\bigg)\bigg]\tau\cdot\nr,\label{eq:uGI2}
\end{align}
and
    \begin{align}
        \Fou(u\cdot v)
        &=\Big(\pPr\times\Bv\Big)\sinc\Pcr\cdot\Big(\pPr\times\Bv\Big)\sinc'\Pcr\Fou(v)\nonumber\\
        &=\Fou\bigg[\frac{1}{\hbar^2}(\sv\times\Bv)\sinc\scrbalt        \cdot(\sv\times\Bv)\sinc'\bigg(\frac{\rmi}{2}\prl\cdot \sv\bigg) v\bigg]\nonumber\\
        &=\Fou\bigg\{\frac{1}{\hbar^2}\frac{1}{2}\int_{-1}^1\rmd\tau\bigg[\sv\times\Bv\bigg(\rv+\frac{\sv\tau}{2}\bigg)\bigg]\cdot\bigg(-\frac{\rmi}{2}\bigg)\int_{-1}^1\rmd\eta\bigg[\sv\times\Bv\bigg(\rv+\frac{\sv\eta}{2}\bigg)\bigg]\eta v\bigg\}\nonumber\\
        \Rightarrow u(\sv)&=\frac{1}{4\rmi\hbar^2}\int_{-1}^1\rmd\tau\bigg[\sv\times\Bv\bigg(\rv+\frac{\sv\tau}{2}\bigg)\bigg]\cdot\int_{-1}^1\rmd\eta\bigg[\sv\times\Bv\bigg(\rv+\frac{\sv\eta}{2}\bigg)\bigg]\eta.\label{eq:uGI3}
    \end{align}
Applying the convolution theorem together with \eqref{eq:uGI1}--\eqref{eq:uGI3} to \eqref{eq:WEstrong} yields 
\begin{theorem}[Weak formulation of the GIWE]
    \label{weakGIWE}
    Let $\Trho_0$ be the density matrix at $t=0$ in an EM field $(\Ev, \Bv)$. For the Stratonovich function with the initial condition
    \begin{align}
        \Fws^0(\Pv,\rv)=C\int\rmd\sv\expstr
        \Trho_0(\sv,\rv),
    \end{align}
    the evolution of $\Fws$ is given by
    \begin{align}
        \PD{\Fws}{t}+\frac{\Pv}{m}\cdot\nr\Fws=&\int\rmd\Pv'L_\rmw(\Pv,\Pv-\Pv',\rv)\Fws(\Pv',\rv)
        +\int\rmd\Pv'{\bf M}_{\rmw 1}(\Pv-\Pv',\rv)\cdot\nr\Fws(\Pv',\rv)
        \nonumber\\
        &+\int\rmd\Pv'M_{\rmw 2}(\Pv-\Pv',\rv)\Fws(\Pv',\rv),
        \label{eq:GIWEweak}
    \end{align}
    where
    \begin{align}
        L_\rmw(\Pv,\Pv',\rv)&:=-\frac{1}{2\rmi\hbar}C\int\rmd\sv\exp{\frac{1}{\rmi\hbar}\sv\cdot\Pv'}\int_{-1}^1\rmd\tau\Fv\Big(\Pv,\rv+\frac{\sv\tau}{2}\Big)\cdot\sv,\label{eq:Lw}\\
        {\bf M}_{\rmw 1}(\Pv,\rv)&:=-\frac{q}{4m}C\int\rmd\sv\exp{\frac{1}{\rmi\hbar}\sv\cdot\Pv}\int_{-1}^1\rmd\tau\Big[\sv\times\Bv\Big(\rv+\frac{\sv\tau}{2}\Big)\Big]\tau,\\
        M_{\rmw 2}(\Pv,\rv)&:=\frac{q^2}{8\rmi\hbar m}C\int\rmd\sv\exp{\frac{1}{\rmi\hbar}\sv\cdot\Pv}\int_{-1}^1\rmd\tau\Big[\sv\times\Bv\Big(\rv+\frac{\sv\tau}{2}\Big)\Big]\cdot
        \int_{-1}^1\rmd\eta\Big[\sv\times\Bv\Big(\rv+\frac{\sv\eta}{2}\Big)\Big]\eta.\label{eq:Mw2}
    \end{align}
\end{theorem}
As the functions \eqref{eq:Lw}--\eqref{eq:Mw2} are defined as Fourier transforms of the fields, a sufficient condition for their existence is that the line integrals in $L_\rmw$ and $\Mv_{\rmw1}$ are $L^1(\R^3)$, and in $M_{\rmw 2}$ are $L^2(\R^3)$.
This poses different requirements to the EM field: The strong formulation imposes strong regularity conditions (smoothness), but does not constrain the asymptotic behavior as the equation is local. In contrast, the integrability of the weak formulation relaxes the regularity requirements but introduces constraints to the asymptotic behavior. When we consider distributions for the EM field, like in the case of point charges, the weak formulation can easily be evaluated in the integrals. This is also possible for the strong formulation by using weak derivatives, which introduces integrals but preserves locality.


Theorem \ref{weakGIWE} is the same formulation as the one derived in \cite{PRBMagnetic}, which also used the WS transform but started from the von Neumann equation. 
Thus, we can conclude the relations between the different formalisms: Starting from the Schr\"{o}dinger equation, the von Neumann equation is derived. From there, the path splits into two, where one directly gives us the weak formulation of the GIWE. 
The other path leads to Moyal's formulation of the Wigner equation, where inserting the Hamiltonian for a charged particle into an EM field yields the strong formulation of the GDWE. 
The concatenation of the inverse Wigner and WS transform converts this equation into the strong formulation of the GIWE. The cycle is closed by transforming the strong into the weak formulation.

The conservation of particles is an important property in quantum mechanics. The continuity equation is crucial to show that no particle can be destroyed or created from nothing. In the next section, we show the relevance of $\Fws$ for the conservation of particles in EM fields.
\subsection{Conservation law for the probability current in EM fields}
According to wave mechanics, the probability current of a charged particle in an EM field is given as 
\begin{equation}
    \jv:=-\frac{\rmi\hbar}{2m}[\psi^*\nabla\psi-\psi\nabla\psi^*]-\frac{q}{m}\psi^*\psi\Av.
\end{equation}
Since the corresponding expression for the current density in gauge-dependent Wigner mechanics is given as $\jv(\rv)=\int m^{-1}\pv\fw(\pv,\rv)\d\pv$, it is natural to assume that this holds also in the gauge-invariant case, using $\Fws$. Indeed, we can prove that
\begin{equation}
    \jv(\rv)=\int\d\Pv\frac{\Pv}{m}\Fws(\Pv,\rv).
\end{equation}
To show this, we need to calculate the derivative of the line integral of $\Av$ at $\sv=\textbf{0}$, which we will apply later in \eqref{eq:probCurrent}. Here we use the $\sinc$ representation like in \eqref{eq:fssinc}: 
\begin{align}
    \ns \bigg[\sv\cdot\sinc\bigg(\frac{\rmi\sv}{2}\cdot\nr\bigg)\Av(\rv)\bigg]\bigg|_{\sv=\textbf{0}}&=\ns \bigg[\sv\cdot\Av(\rv)+\frac{1}{24}(\sv\cdot\nr)^2\sv\cdot\Av(\rv)+\dots\bigg]\bigg|_{\sv=\textbf{0}}\nonumber\\
    &=\Av(\rv),
\end{align}
We also calculate
\begin{align}
    \ns\Trho(\sv,\rv)|_{\sv=\textbf{0}}&=\ns\Big[\psi\Big(\rv+\frac{\sv}{2}\Big)\psi^*\Big(\rv-\frac{\sv}{2}\Big)\Big]\Big|_{\sv=\textbf{0}}\nonumber\\
    &=\frac{1}{2}[\nabla\psi(\rv)\psi^*(\rv)-\psi(\rv)\nabla\psi^*(\rv)],
\end{align}
which gives us
\begin{align}
    \int\d\Pv\frac{\Pv}{m}\Fws(\Pv,\rv)&=C\int\d\Pv\int\d\sv\frac{\rmi\hbar}{m}\ns\exp{\frac{1}{\rmi\hbar}\sv\cdot\Pv}\expsq{\frac{q}{2\rmi\hbar}\sv\cdot\int_{-1}^1\d\tau\Av\Big(\rv+\frac{\sv\tau}{2}\Big)}\Trho(\sv,\rv)\nonumber\\
    &=-C\int\d\Pv\int\d\sv\frac{\rmi\hbar}{m}\exp{\frac{1}{\rmi\hbar}\sv\cdot\Pv}\ns\bigg\{\expsq{\frac{q}{2\rmi\hbar}\sv\cdot\int_{-1}^1\d\tau\Av\Big(\rv+\frac{\sv\tau}{2}\Big)}\Trho(\sv,\rv)\bigg\}\nonumber\\   
    &=-\int\d\sv\delta(\sv)\frac{\rmi\hbar}{m}\expsq{\frac{q}{2\rmi\hbar}\sv\cdot\int_{-1}^1\d\tau\Av\Big(\rv+\frac{\sv\tau}{2}\Big)}\nonumber\\
    &\quad\qquad\bigg\{\frac{q}{\rmi\hbar}\ns \bigg[\sv\cdot\sinc\bigg(\frac{\rmi\sv}{2}\cdot\nr\bigg)\Av(\rv)\bigg]+\ns\Trho(\sv,\rv)\bigg\}\nonumber\\
    &=-\frac{\rmi\hbar}{2m}[\nabla\psi(\rv)\psi^*(\rv)-\psi(\rv)\nabla\psi^*(\rv)]-\frac{q}{m}\Av(\rv)\nonumber\\
    &=\jv(\rv).
    \label{eq:probCurrent}
\end{align}
We also find that the probability density can be calculated as
\begin{align}
    P(\rv)=\int\d\Pv\Fws(\Pv,\rv).
\end{align}
Thus, the continuity equation can be rewritten as
\begin{equation}
    \PD{P}{t}(\rv)+\nr\cdot\jv(\rv)=0\Leftrightarrow\int\d\Pv\bigg[\PD{\Fws}{t}(\Pv,\rv)+\frac{\Pv}{m}\cdot\nr\Fws(\Pv,\rv)\bigg]=0,
\end{equation}
where the left-hand side of the GIWE can immediately be identified inside the integral. This implies that integrating the quantum correction terms over $\Pv$ yields zero, and the integration of the whole GIWE directly yields the continuity equation. This shows that the gauge-invariant phase-space description respects probability conservation and thus consistently describes observable density and current quantities.

\section{Conclusion}
\label{conclusion}
The gauge-independent Wigner equation for a single electron in an electromagnetic field is derived for a strong formulation, which uses expansions of differential operators, and a weak formulation, which uses convolutions. The derivations are based on Moyal's formulation of the Wigner equation. This formulation introduces the sine function of the differential operators acting on the Wigner function and the Hamiltonian. This yields a strong formulation for the gauge-dependent Wigner equation, which can be transformed into a weak formulation using a series expansion. The strong formulation can then be transformed to make the equation gauge-invariant. For this purpose, a variation of the Wigner function must be introduced. The Weyl-Stratonovich transform defines a new function by replacing the canonical momentum with the kinetic momentum and changing the exponent of the Weyl transform. In this way, a gauge-invariant function, called the Stratonovich function, is introduced. For the transformation, an inverse Weyl transform is applied to the Wigner function, which yields the density matrix. Then the Weyl-Stratonovich transform gives us the Stratonovich function. Using this approach for the whole equation yields an evolution equation. To show its gauge-invariance, all gauge variables $\Av$ and $\phi$ must be replaced by the field variables $\Ev$ and $\Bv$. This finally gives us the strong formulation of the gauge-invariant Wigner equation, which is then transformed into the weak formulation using the same approach as in the gauge-dependent case. Finally, it is shown that the integration of the Stratonovich function over the momentum space yields the continuity equation, which states that the total probability is preserved during the evolution.

\begin{acknowledgments}
This research was funded in whole by the Austrian Science Fund (FWF) P37080-N.
\end{acknowledgments}

\appendix
\section{Weyl-Stratonovich transform}
\subsection{Gauge-dependence of $\Fw$}
\label{eq:gdFw}
\begin{align}
    \Fw'(\Pv,\rv,t)
    &=C\int\rmd\sv\expcu{\frac{1}{\rmi\hbar}\sv\cdot[\Pv+q\Av'(\rv)]}\Trho'(\sv,\rv)\nonumber\\
    &=C\int\rmd\sv\expcu{\frac{1}{\rmi\hbar}\sv\cdot[\Pv+q\Av(\rv)+q\nabla\chi(\rv)]}\nonumber\\
    &\quad\expcu{-\frac{1}{\rmi\hbar} q\Big[\chi\Big(\rv+\frac\sv 2\Big)-\chi\Big(\rv-\frac\sv 2\Big)\Big]}\Trho(\sv,\rv)\nonumber\\
    &\neq C\int\rmd\sv\expcu{\frac{1}{\rmi\hbar}\sv\cdot[\Pv+q\Av(\rv)]}\Trho(\sv,\rv)\nonumber\\
    &=\Fw(\Pv,\rv,t).
\end{align}
\subsection{Proof for Lemma~\ref{lemmaGIFs}}
\label{proofGIFs}
\begin{proof}
    First, we transform
    \begin{align*}
        \chi\Big(\rv+\frac{\sv}{2}\Big)-\chi\Big(\rv-\frac{\sv}{2}\Big)&=\int_{-1}^1\rmd\tau\DD{\chi}{\tau}\Big(\rv+\frac{\sv\tau}{2}\Big)\\
        &=\int_{-1}^1\rmd\tau\frac{\sv}{2}\cdot\nabla\chi\Big(\rv+\frac{\sv\tau}{2}\Big),
    \end{align*}
    which is used to show that
    \begin{align*}
        &\Fws'(\Pv,\rv)\\
        &= C\int\rmd\sv\expcu{\frac{1}{\rmi\hbar}\sv\cdot\left[\Pv+\frac{q}{2}\int_{-1}^1\rmd\tau\Av'\left(\rv+\frac{\sv\tau}{2}\right)\right]}\Trho'(\sv,\rv)\\
        &= C\int\rmd\sv\,{\rm exp}\bigg\{{\frac{1}{\rmi\hbar}\sv\cdot\bigg[\Pv+\frac{q}{2}\int_{-1}^1\rmd\tau\Av\left(\rv+\frac{\sv\tau}{2}\right)}\\
        &~{+\nabla\chi\left(\rv+\frac{\sv\tau}{2}\right)\bigg]}\bigg\}\expcu{-\frac{q}{\rmi\hbar} \left[\chi\left(\rv+\frac{\sv}{2}\right)-\chi\left(\rv-\frac{\sv}{2}\right)\right]}\Trho(\sv,\rv)\\
        &= C\int\rmd\sv\,{\rm exp}\bigg\{{\frac{1}{\rmi\hbar}\sv\cdot\bigg[\Pv+\frac{q}{2}\int_{-1}^1\rmd\tau\Av\left(\rv+\frac{\sv\tau}{2}\right)}\\
        &~{+\nabla\chi\left(\rv+\frac{\sv\tau}{2}\right)\bigg]}\bigg\}\expcu{-\frac{q}{2\rmi\hbar}\int_{-1}^1\rmd\tau\sv\cdot\nabla\chi\left(\rv+\frac{\sv\tau}{2}\right)}\Trho(\sv,\rv)\\
        &= C\int\rmd\sv\expstr\Trho(\sv,\rv)\\
        &=\Fws(\Pv,\rv).
    \end{align*} 
\end{proof}
\subsection{Properties of $\SW$: Derivatives}
\label{Cderivative}
\begin{proof}
    We can replace the line integral in the exponent of $\Fws$ by $\sinc$ using \eqref{eq:sincInt}, so that 
\begin{align}
    \Fws(\Pv,\rv)=
    C\int\rmd\sv\expstralt\Trho(\sv,\rv).
    \label{eq:fssinc}
\end{align}
Note that the nabla operator in the $\sinc$-function only acts on $\Av$.
    The proof for $\partial/\partial t$ and $\nr$ is the same. Therefore, we write $\alpha$, which could stand for $x,y,z$ or $t$.
    \begin{align*}
        \PD{\Fws}{\alpha}=&\int\rmd\sv\expcu{\frac{1}{\rmi\hbar}\sv\cdot[\Pv+q\sinc\scr\Av(\rv,t)]}\PD{\Trho}{\alpha}(\sv,\rv)\\
        &+\int\rmd\sv\PD{}{\alpha}\expcu{\frac{1}{\rmi\hbar}\sv\cdot[\Pv+q\sinc\scr\Av(\rv,t)]}\Trho(\sv,\rv)\\
        =&\,\SW\bigg(\PD{\fw}{\alpha}\bigg)+\frac{q}{\rmi\hbar}\int\rmd\sv\sinc\scrb\bigg[\sv\cdot\PD{\Av}{\alpha}(\rv,t)\bigg]\\
        &\expcu{\frac{1}{\rmi\hbar}\sv\cdot\bigg[\Pv+q\sinc\scr\Av(\rv,t)\bigg]}\Trho(\sv,\rv)\\
        =&\,\SW\bigg(\PD{\fw}{\alpha}\bigg)+q\bigg[\pPr\cdot\PD{\Av}{\alpha}(\rv,t)\bigg]\sinc\Pcr\\
        &\int\rmd\sv\expcu{\frac{1}{\rmi\hbar}\sv\cdot\bigg[\Pv+q\sinc\scr\Av(\rv,t)\bigg]}\Trho(\sv,\rv)\\
        =&\,\SW\bigg(\PD{\fw}{\alpha}\bigg)+q\bigg[\pPr\cdot\PD{\Av}{\alpha}(\rv,t)\bigg]\sinc\Pcr\Fws,
    \end{align*}
    where we have used Lemma~\ref{stonp}.
    The same approach cannot be used for $\np$ as the exponential function of $\fw$ depends on $\pv$. Thus, we show
        \begin{align*}
            &\nP\Fws\\
            &=\nP\int\rmd\sv\expcu{\frac{1}{\rmi\hbar}\sv\cdot\bigg[\Pv+q\sinc\scr\Av(\rv,t)\bigg]}\Trho(\sv,\rv)\\
            &=\frac{1}{\rmi\hbar}\int\rmd\sv\expcu{\frac{1}{\rmi\hbar}\sv\cdot\bigg[\Pv+q\sinc\scr\Av(\rv,t)\bigg]}\sv \Trho(\sv,\rv)\\
            &=\SW\bigg[\frac{1}{\rmi\hbar}\int\rmd\sv\exp{\frac{1}{\rmi\hbar}\sv\cdot\pv}\sv \Trho(\sv,\rv)\bigg]\\
            &=\SW\bigg[\np\int\rmd\sv\exp{\frac{1}{\rmi\hbar}\sv\cdot\pv}\Trho(\sv,\rv)\bigg]\\
            &=\SW[\np\fw].
        \end{align*}
\end{proof}
\subsection{Properties of $\SW$: $\pv\fw$}
\label{CP}
First, we calculate the inverse Weyl transformation of $\pv\fw$, using $C\int\rmd\pv e^{\rmi/\hbar(\sv-\sv')\cdot\pv}=\delta(\sv-\sv')$.
        \begin{align*}
            \Wig^{-1}(\pv\fw)&=C\int\rmd\pv\exp{-\frac{1}{\rmi\hbar}\sv\cdot\pv}\pv\int\rmd\sv'\exp{\frac{1}{\rmi\hbar}\sv'\cdot\pv} \Trho(\sv',\rv)\\
            &=\rmi\hbar\int\rmd\sv'  C\int\rmd\pv \nabla_{\sv'}\Big(e^{-\frac{1}{\rmi\hbar}(\sv-\sv')\cdot\pv}\Big)\Trho(\sv',\rv)\\
            &=-\rmi\hbar\int\rmd\sv' C\int\rmd\pv e^{-\frac{1}{\rmi\hbar}(\sv-\sv')\cdot\pv}\nabla_{\sv'}\Trho(\sv',\rv)\\
            &=-\rmi\hbar\int\rmd\sv'\delta(\sv-\sv')\nabla_{\sv'}\Trho(\sv',\rv)\\
            &=-\rmi\hbar\nabla_{\sv}\Trho(\sv,\rv).
        \end{align*}
        Next, with the help of Lemma~\ref{stonp}, we apply the WS transform, which yields
            \begin{align*}
                \SW(\pv\fw)=&\,\Str\big(-\rmi\hbar\nabla_{\sv}\Trho(\sv,\rv)\big)\\
                =&-\rmi\hbar C\int\rmd\sv\expcu{\frac{1}{\rmi\hbar}\sv\cdot\bigg[\Pv+q\sinc\scr\Av(\rv)\bigg]}\nabla_\sv \Trho(\sv,\rv)\\
                =&\,\rmi\hbar C\int\rmd\sv\nabla_{\sv} \expcu{\frac{1}{\rmi\hbar}\sv\cdot\bigg[\Pv+q\sinc\scr\Av(\rv)\bigg]}\Trho(\sv,\rv)\\
                =&\, C\int\rmd\sv\bigg[\Pv+q\sinc\scrb\Av(\rv)+\frac{\rmi q}{2}\sinc'\scrb\nr(\sv\cdot\Av)\bigg]
                \nonumber\\&
                \expcu{\frac{1}{\rmi\hbar}\sv\cdot\bigg[\Pv+q\sinc\scr\Av(\rv)\bigg]}\Trho(\sv,\rv)\\
                =&\, C\int\rmd\sv\bigg[\Pv+q\Av(\rv)\sinc\bigg(-\frac{\hbar}{2}\prl\cdot\pPr\bigg)-\frac{q\hbar}{2}\nr(\pPr\cdot\Av)\sinc'\bigg(-\frac{\hbar}{2}\prl\cdot\pPr\bigg)\bigg]\nonumber\\
                &\expcu{\frac{1}{\rmi\hbar}\sv\cdot\bigg[\Pv+q\sinc\scr\Av(\rv)\bigg]}\Trho(\sv,\rv)\\
                =&\bigg[\Pv+q\Av(\rv)\sinc\bigg(\frac{\hbar}{2}\prl\cdot\pPr\bigg)+\frac{q\hbar}{2}\nr(\pPr\cdot\Av)\sinc'\bigg(\frac{\hbar}{2}\prl\cdot\pPr\bigg)\bigg]\Fws(\sv,\rv)
            \end{align*}
    \section{Gauge-invariant WE}
    \subsection{Derivation of the Strong formulation}
    \label{DevSF}
    We start with the strong formulation of the GDWE and expand the brackets:
        \begin{align}
            \PD{\fw}{t}+\frac{2q}{\hbar m}\pv\cdot\Av\sin\bigg(\frac{\hbar}{2}\prl\cdot\ppr\bigg)\fw-\frac{q^2}{\hbar m}\Av^2\sin\bigg(\frac{\hbar}{2}\prl\cdot\ppr\bigg)\fw-\frac{2q}{\hbar}\phi\sin\bigg(\frac{\hbar}{2}\prl\cdot\ppr\bigg)\fw\nonumber\\
            +\frac{\pv}{m}\cdot\nr\fw-\frac{q}{m}\Av\cdot\cos\bigg(\frac{\hbar}{2}\prl\cdot\ppr\bigg)\nr\fw=0.
        \end{align}
    Then we transform each term as described in step 1. The terms, for which we will use the identity \eqref{eq:vtp} from step 2, and $\Ev=-\nr\phi-\partial\Av/\partial t$, are labeled with Roman numbers to identify which of them are combined. We also apply the identities \eqref{eq:sinSinc} and \eqref{eq:cosSinc} of step 4:
    \begin{itemize}
        \item The transformation of the time derivative yields
        \begin{equation}
            \SW\bigg(\PD{\fw}{t}\bigg)\overset{\eqref{eq:fwdt}}{=}\PD{\Fws}{t}-q\PD{\Av}{t}\cdot\sinc\Pcr\nP\Fws^{\mathrm{I.a}},
        \end{equation}
        \item By using the multiplicativity of the transformation, the term containing $\pv\cdot\Av$ transforms as
        \begin{align}
            &\SW\bigg[\frac{2q}{\hbar m}\pv\cdot\Av\sin\bigg(\frac{\hbar}{2}\prl\cdot\ppr\bigg)\fw\bigg]\nonumber\\
            &=\frac{2q}{\hbar m}\Av\sin\bigg(\frac{\hbar}{2}\prl\cdot\ppr'\bigg)\cdot\SW(\pv\fw')\nonumber\\
            &\overset{\eqref{eq:pfw}}{=}\frac{2q}{\hbar m}\bigg[\Pv+\frac{q\hbar}{2}\nr\Big(\pPr\cdot\Av\Big)\sinc'\Pcr+q\Av\sinc\Pcr\bigg]\cdot\Av\sin\bigg(\frac{\hbar}{2}\prl\cdot\pPr\bigg)\Fws\nonumber\\
            &\overset{\eqref{eq:sinSinc}}{=}\frac{q}{m}\Pv\cdot\Av\Pcrb\sinc\bigg(\frac{\hbar}{2}\prl\cdot\pPr\bigg)\Fws^{\mathrm{II.a}}
            \nonumber\\&
            \quad+\frac{q^2}{m}\Av\sinc\Pcr\cdot\Av\Pcrb\sinc\bigg(\frac{\hbar}{2}\prl\cdot\pPr\bigg)\Fws^{\mathrm{III.a}}\nonumber\\
            &\quad+\frac{q^2\hbar}{2m}\nr\Big(\pPr\cdot\Av\Big)\sinc'\Pcr\cdot\Av\Pcrb\sinc\bigg(\frac{\hbar}{2}\prl\cdot\pPr\bigg)\Fws^{\mathrm{IV.a}}.
        \end{align}
        \item For the term containing $\Av^2$ we use \eqref{eq:Asquared} from step 3, which gives us
        \begin{align}
            &\SW\bigg[-\frac{q^2}{\hbar m}\Av^2\sin\bigg(\frac{\hbar}{2}\prl\cdot\ppr\bigg)\fw\bigg]\overset{\eqref{eq:Asquared}}{=}-\frac{2q^2}{\hbar m}\bigg[\Av\sin\Pcr\bigg]\cdot\bigg[\Av\cos\Pcr\bigg]\Fws\nonumber\\
            &\quad\overset{\eqref{eq:sinSinc},\eqref{eq:cosSinc}}{=}-\frac{q^2}{m}\Av\Pcrb\sinc\Pcr
            \nonumber\\&
            \cdot\bigg[\Av\sinc\Pcr+\frac{\hbar}{2}\Av\Pcrb\sinc'\Pcr\bigg]\Fws\nonumber\\
            &\quad=-\frac{q^2}{m}\Av\Pcrb\sinc\Pcr\cdot\Av\sinc\Pcr\Fws^{\mathrm{III.b}}\nonumber\\
            &\qquad\quad-\frac{q^2\hbar}{2m}\Av\Pcrb\sinc\Pcr\cdot\Av\Pcrb\sinc'\Pcr\Fws^{\mathrm{IV.b}}.
        \end{align}
        \item The term with the scalar potential $\phi$ becomes
        \begin{equation}
            \SW\bigg[-\frac{2q}{\hbar}\phi\sin\bigg(\frac{\hbar}{2}\prl\cdot\ppr\bigg)\fw\bigg]\overset{\eqref{eq:sinSinc}}{=}-q\nr\phi\cdot\sinc\Pcr\nP\Fws^{\mathrm{I.b}}.
        \end{equation}
        \item The term containing $\nabla_r f_w$ represents a diffusion term and transforms as
        \begin{align}
            &\SW\bigg\{\bigg[\frac{\pv}{m}-\frac{q}{m}\Av\cos\Pcrb\bigg]\cdot\nr\fw\bigg\}\overset{\eqref{eq:pfw},\eqref{eq:cosSinc}}{=}\bigg[\frac{\Pv}{m}+\frac{q\hbar}{2m}\nr\Big(\pPr\cdot\Av\Big)\sinc'\Pcr
            \nonumber\\ & 
            +\hcancel{\frac{q}{m}\Av\sinc\Pcr}\nonumber\\
            &\qquad-\hcancel{\frac{q}{m}\Av\sinc\Pcr}-\frac{q\hbar}{2m}\Av\Pcrb\sinc'\Pcr\bigg]\cdot\SW(\nr\fw)\nonumber\\
            &\quad\overset{\eqref{eq:fwdr},\eqref{eq:vtp}}{=}\bigg[\frac{\Pv}{m}+\frac{q\hbar}{2m}\Big(\pPr\times\Bv\Big)\sinc'\Pcr\bigg]\cdot\bigg[\nr\Fws-q\nr(\pPr\cdot\Av)\sinc\Pcr\Fws\bigg]\nonumber\\
            &\quad=\frac{\Pv}{m}\cdot\nr\Fws+\frac{q\hbar}{2m}\Big(\pPr\times\Bv\Big)\sinc'\Pcr\cdot\nr\Fws-\frac{q}{m}\Pv\cdot\nr(\pPr\cdot\Av)\sinc\Pcr\Fws^{\mathrm{II.b}}\nonumber\\
            &\qquad-\frac{q^2\hbar}{2m}\Big(\pPr\times\Bv\Big)\sinc'\Pcr\cdot\nr(\pPr\cdot\Av)\sinc\Pcr\Fws^{\mathrm{IV.c}}\label{eq:diffT}.
        \end{align}
    \end{itemize}
    Now it is time to sum up all terms as described in step 5. The two terms labeled with $\mathrm{III}$ add up to zero. For $\mathrm{I}$ we use $\Ev=-\nabla\phi-\partial\Av/\partial t$, which yields
    \begin{align}
        \mathrm{I}&=\mathrm{I.a}+\mathrm{I.b}\nonumber\\
        &=-q\PD{\Av}{t}\cdot\sinc\Pcr\nP\Fws-q\nr\phi\cdot\sinc\Pcr\nP\Fws\nonumber\\
        &=q\Ev\cdot\sinc\Pcr\nP\Fws.
    \end{align}
    Using \eqref{eq:vtp} and the scalar triple product formula transforms $\mathrm{II}$ as
    \begin{align}
        \mathrm{II}&=\mathrm{II.a}+\mathrm{II.b}\nonumber\\
        &=\frac{q}{m}\Pv\cdot\Av\Pcrb\sinc\bigg(\frac{\hbar}{2}\prl\cdot\pPr\bigg)\Fws-\frac{q}{m}\Pv\cdot\nr(\pPr\cdot\Av)\sinc\Pcr\Fws\nonumber\\
        &=-\frac{q}{m}\Pv\cdot\Big(\pPr\times\Bv\Big)\sinc\bigg(\frac{\hbar}{2}\prl\cdot\pPr\bigg)\Fws\nonumber\\
        &=\frac{q}{m}(\Pv\times\Bv)\cdot\pPr\sinc\bigg(\frac{\hbar}{2}\prl\cdot\pPr\bigg)\Fws.
    \end{align}
    Also for $\mathrm{IV}$ \eqref{eq:vtp} is used to evaluate
    \begin{align}
        \mathrm{IV}&=\mathrm{IV.a}+\mathrm{IV.b}+\mathrm{IV.c}\nonumber\\
        &=\frac{q^2\hbar}{2m}\nr\Big(\pPr\cdot\Av\Big)\sinc'\Pcr\cdot\Av\Pcrb\sinc\bigg(\frac{\hbar}{2}\prl\cdot\pPr\bigg)\Fws\nonumber\\
        &\quad-\frac{q^2\hbar}{2m}\Av\Pcrb\sinc\Pcr\cdot\Av\Pcrb\sinc'\Pcr\Fws\nonumber\\
        &\quad-\frac{q^2\hbar}{2m}\Big(\pPr\times\Bv\Big)\sinc'\Pcr\cdot\nr(\pPr\cdot\Av)\sinc\Pcr\Fws\nonumber\\
        &=\frac{q^2\hbar}{2m}\Big(\pPr\times\Bv\Big)\sinc'\Pcr\cdot\Av\Pcrb\sinc\bigg(\frac{\hbar}{2}\prl\cdot\pPr\bigg)\Fws\nonumber\\
        &\quad-\frac{q^2\hbar}{2m}\Big(\pPr\times\Bv\Big)\sinc'\Pcr\cdot\nr(\pPr\cdot\Av)\sinc\Pcr\Fws\nonumber\\
        &=-\frac{q^2\hbar}{2m}\Big(\pPr\times\Bv\Big)\sinc'\Pcr\cdot\Big(\pPr\times\Bv\Big)\sinc\bigg(\frac{\hbar}{2}\prl\cdot\pPr\bigg)\Fws.
    \end{align}
    Adding up all terms gives us
    \begin{align}
        &\PD{\Fws}{t}+\frac{\Pv}{m}\cdot\nr\Fws+\frac{q\hbar}{2m}\Big(\pPr\times\Bv\Big)\sinc'\Pcr\cdot\nr\Fws+\mathrm{I}+\mathrm{II}+\mathrm{IV}=0\nonumber\\
        \Leftrightarrow&\,\PD{\Fws}{t}+\frac{\Pv}{m}\cdot\nr\Fws+\frac{q\hbar}{2m}\Big(\pPr\times\Bv\Big)\sinc'\Pcr\cdot\nr\Fws+q\Ev\cdot\sinc\Pcr\nP\Fws\nonumber\\
        &+\frac{q}{m}(\Pv\times\Bv)\cdot\pPr\sinc\bigg(\frac{\hbar}{2}\prl\cdot\pPr\bigg)\Fws-\frac{q^2\hbar}{2m}\Big(\pPr\times\Bv\Big)\sinc'\Pcr\cdot\Big(\pPr\times\Bv\Big)
        \nonumber\\&
        \sinc\bigg(\frac{\hbar}{2}\prl\cdot\pPr\bigg)\Fws=0.
    \end{align}
    To simplify the remaining terms according to step 6, we factor out the $\sinc'$-expression and use $\Fv=q\Ev+qm^{-1}\Pv\times\Bv$, yielding
    \begin{align}
        &\PD{\Fws}{t}+\frac{\Pv}{m}\cdot\nr\Fws+\Fv\cdot\sinc\Pcr\nP\Fws\nonumber\\
        &\quad=-\frac{q\hbar}{2m}\Big(\pPr\times\Bv\Big)\sinc'\Pcr
        \cdot\bigg[\prr-q\Big(\pPr\times\Bv\Big)\sinc\Pcr\bigg]\Fws.
    \end{align}
\section{GDWE}
\subsection{Proof for Lemma~\ref{stonp}}
\label{proofStonp}
\begin{proof}
    As $g$ is analytic, we can write it as $g(x)=\sum_{n=0}^\infty a_nx^n$. We can also exchange the limit of the expansion with the integral, due to its uniform convergence. For a pseudo-differential operator of a polynomial $p(D)=\sum_\alpha a_\alpha D^\alpha,D^\alpha=(-\rmi\hbar\partial_1)^{\alpha_1}\cdots(-\rmi\hbar\partial_n)^{\alpha_n}$ holds
    \begin{equation*}
        \Fou[p(\sv)f(\sv)]=p(D)\Fou[f(\sv)].
    \end{equation*}
    This is used to show that
    \begin{align*}
        \Fou\Big[g\Big(\rmi\prl\cdot\sv\Big)f(\sv)\Big]\\
        =\sqrt{C}\int\rmd\sv g\Big(\rmi\prl\cdot\sv\Big)\exp{\frac{1}{\rmi\hbar}\sv\cdot\pv}f(\sv)\\
        =\sqrt{C}\int\rmd\sv\sum_{n=0}^\infty a_n\Big(\rmi\prl\cdot\sv\Big)^n\exp{\frac{1}{\rmi\hbar}\sv\cdot\pv}f(\sv)\\
        =\sqrt{C}\lim_{N\rightarrow\infty}\int\rmd\sv\sum_{n=0}^N a_n\Big(\rmi\prl\cdot\sv\Big)^n\exp{\frac{1}{\rmi\hbar}\sv\cdot\pv}f(\sv)\\
        =\sqrt{C}\int\rmd\sv\sum_{n=0}^\infty a_n\Big(-\hbar\prl\cdot\ppr\Big)^n\exp{\frac{1}{\rmi\hbar}\sv\cdot\pv}f(\sv)\\
        =\sqrt{C}g\Big(-\hbar\prl\cdot\ppr\Big)\Fou(f)(\pv).
    \end{align*}
\end{proof}
\subsection{Proof for Lemma~\ref{trigLemma}}
\label{trigLemmaProof}
\begin{proof}
    Using a Taylor expansion for $\tau$ yields
    \begin{align*}
        \int_{-1}^1\rmd\tau g\bigg(\rv+\frac{\sv\tau}{2}\bigg)=&\,\int_{-1}^1\rmd\tau\sum_{n=0}^\infty\frac{1}{n!}\bigg(\frac{\sv\tau}{2}\cdot\nabla\bigg)^n g(\rv)\\
        =&\,\sum_{n=0}^\infty\frac{1}{n!}\frac{1}{2^n}(\sv\cdot\nabla)^n g(\rv)\frac{1}{n+1}\big(1^{n+1}-(-1)^{n+1}\big)\\
        =&\,2\sum_{n=0}^\infty\frac{1}{(2n+1)!}\frac{1}{2^{2n}}(\sv\cdot\nabla)^{2n} g(\rv)\\
        =&\,2\sinc\bigg(\frac{\rmi\sv}{2}\cdot\nabla\bigg) g(\rv).\\
    \end{align*}
    Similarly, we can show that
    \begin{equation*}
        \cos\bigg(\frac{\rmi\sv}{2}\cdot\nabla\bigg) g(\rv)=\frac{1}{2}\bigg[g\bigg(\rv+\frac{\sv}{2}\bigg)+g\bigg(\rv-\frac{\sv}{2}\bigg)\bigg],
    \end{equation*}
    which we use together with \eqref{eq:cosSinc} and partial integration to show that
    \begin{align*}
        g(\rv)\scrbalt\sinc'\scrbalt\nonumber\\
        =g(\rv)\cos\scrbalt-g(\rv)\sinc\scrbalt\nonumber\\
        =\frac{1}{2}\bigg[g\bigg(\rv+\frac{\sv}{2}\bigg)+g\bigg(\rv-\frac{\sv}{2}\bigg)\bigg]-\frac{1}{2}\int_{-1}^1\rmd\tau g\bigg(\rv+\frac{\sv\tau}{2}\bigg)\nonumber\\
        =\frac{1}{2}\bigg[g\bigg(\rv+\frac{\sv\tau}{2}\bigg)\tau\bigg]_{\tau=-1}^1-\frac{1}{2}\int_{-1}^1\rmd\tau g\bigg(\rv+\frac{\sv\tau}{2}\bigg)\nonumber\\
        =\frac{1}{2}\int_{-1}^1\rmd\tau\PD{g}{\tau}\bigg(\rv+\frac{\sv\tau}{2}\bigg)\tau\nonumber\\
        =\frac{1}{2}\int_{-1}^1\rmd\tau\frac{\sv}{2}\cdot\nr g\bigg(\rv+\frac{\sv\tau}{2}\bigg)\tau\nonumber\\
        =-\frac{\rmi}{2}\bigg(\frac{\rmi}{2}\sv\cdot\nr\bigg)\int_{-1}^1\rmd\tau g\bigg(\rv+\frac{\sv\tau}{2}\bigg)\tau\\
        \Rightarrow g(\rv)\sinc'\scrbalt=-\frac{\rmi}{2}\int_{-1}^1\rmd\tau g\bigg(\rv+\frac{\sv\tau}{2}\bigg)\tau.
    \end{align*}
\end{proof}
    \subsection{Connection between $\rho_{\rm w}$ and $\Fws$}
    \label{rhowVSfs}
    The gauge-invariant Wigner operator is defined in \cite{Levanda01} as
\begin{equation}
    \hat{\rho}_\rmw(P,X)=\frac{C}{2\pi}\int\rmd^4y{\rm exp}\bigg[\frac{1}{\rmi\hbar}Py+\frac{q}{\rmi\hbar}\int_{-1/2}^{1/2}y^\mu A_\mu(X+s y)\rmd s\bigg]\hat{\rho}(y,X),
\end{equation}
using index notation $\mu$ with the 4-vectors $P=(\Pv,\epsilon),X=(\rv,t),y=(\sv,\tau),A=(\Av,\phi)$, is connected to $\Fws$ via
\begin{align}
    \int\rmd\epsilon\hat{\rho}_\rmw(P,X)&=\frac{C}{2\pi}\int\rmd\epsilon\int\rmd^4y{\rm exp}\bigg[\frac{1}{\rmi\hbar}Py+\frac{q}{\rmi\hbar}\int_{-1/2}^{1/2}y^\mu A_\mu(X+s y)\rmd s\bigg]\hat{\rho}(y,X)\nonumber\\
    &=\frac{C}{2\pi}\int\rmd\epsilon\int\rmd\tau\int\rmd\sv{\rm exp}\bigg[\frac{1}{\rmi\hbar}(\sv\cdot\Pv+\tau\epsilon)\nonumber\\
    &\quad+\frac{q}{\rmi\hbar}\int_{-1}^{1}\rmd\gamma\sv\cdot\Av\Big(\rv+\frac{\sv\gamma}{2},t+\frac{\tau\gamma}{2}\Big)+\tau\phi\Big(\rv+\frac{\sv\gamma}{2},t+\frac{\tau\gamma}{2}\Big)\bigg]\hat{\rho}\Big(\rv+\frac{\sv}{2},t+\frac{\tau}{2},\rv-\frac{\sv}{2},t-\frac{\tau}{2}\Big)\nonumber\\
    &=C\int\rmd\tau\int\rmd\sv\delta(\tau){\rm exp}\bigg[\frac{1}{\rmi\hbar}\sv\cdot\Pv\nonumber\\
    &\quad+\frac{q}{\rmi\hbar}\int_{-1}^{1}\rmd\gamma\sv\cdot\Av\Big(\rv+\frac{\sv}{2},t+\frac{\tau}{2}\Big)+\tau\phi\Big(\rv+\frac{\sv\gamma}{2},t+\frac{\tau\gamma}{2}\Big)\bigg]\hat{\rho}\Big(\rv+\frac{\sv}{2},t+\frac{\tau}{2},\rv-\frac{\sv}{2},t-\frac{\tau}{2}\Big)\nonumber\\
    &=C\int\rmd\sv{\rm exp}\bigg[\frac{1}{\rmi\hbar}\sv\cdot\Pv
    +\frac{q}{\rmi\hbar}\int_{-1}^{1}\rmd\gamma\sv\cdot\Av\Big(\rv+\frac{\sv\gamma}{2},t\Big)\bigg]\hat{\rho}\Big(\rv+\frac{\sv}{2},t,\rv-\frac{\sv}{2},t\Big)\nonumber\\
    &=\Fws(\Pv,\rv,t),
\end{align}
which shows that $\Fws$ is a marginal distribution of $\hat\rho_\rmw$.

\end{widetext}
\bibliography{apssamp}

\end{document}